\documentstyle[prl,aps]{revtex}

\title{The energy-level statistics in the core of
a vortex in a p-wave superconductor.}

\author{D. A. Ivanov}

\address{Institut f\"ur Theoretische Physik,
ETH-H\"onggerberg, CH-8093 Z\"urich, Switzerland}

\date{November 10, 1999.}

\begin{document}

\maketitle

\begin{abstract}
In the presence of strong disorder, the statistics of
quasiparticle levels in the core
of a vortex in a two-dimensional $p$-wave superconductor
belongs to the universality class $B$
corresponding to the ensemble of orthogonal matrices in
odd dimensions. This novel universality class appears
as a consequence of the $O(2)$ spin symmetry of $p$-wave
pairing. It is preserved in the presence of random disorder,
of electromagnetic vector potential, and of an admixture of the
pairing of opposite chirality in the vortex core, but may be
destroyed by spin-orbit coupling and by Zeeman splitting.
\end{abstract}

\bigskip

The indications of $p$-wave superconductivity in Sr$_2$RuO$_4$
\cite{rutenates} stimulated the study of exotic properties of
$p$-wave superconductors. The order parameter in this compound
is expected to be the same as in the $A$ phase of $^3$He,
$\hat d_\pm({\bf k}) \propto \hat z (k_x \pm ik_y)$. 
The direction of the vector $\hat z$ of the triplet orientation
is fixed by the anisotropy of Sr$_2$RuO$_4$ to be perpendicular
to the Ru-O planes. Because of the strong anisotropy, one may consider
a two-dimensional model as a starting approximation. In two
dimensions, the $p$-wave superconducting gap does not have
nodes, which resembles conventional superconductors.
However, many differences appear in inhomogeneous
setups including boundaries, vortices, and impurities. In particular,
impurities generate bound states with circular currents \cite{Okuno};
similar subgap states appear at the boundary and at
domain walls \cite{Matsumoto}; a single-quantum vortex posesses
a zero-energy state of topological origin \cite{Kopnin,Volovik}.

It is this last property that motivates the present work. The
zero-energy state will be shown to survive a certain class of
disordered perturbations, such as a random potential (modeling
the impurities) or a random electromagnetic vector potential. 
If such a disorder is strong, the quasiparticle
levels in the vortex core mix and, at the energy scale
much smaller than the superconducting gap, they may be described
by a random-matrix ensemble. In the present paper we identify 
the corresponding ensemble with that of orthogonal matrices in
odd number of dimensions (class $B$ in Cartan's 
classification) \cite{Helgason}.
Thus this example completes the list of universality classes
corresponding to the eleven families of symmetric spaces
(see Table 1).
The three Wigner-Dyson universality classes correspond to
$A$, $A$I, and $A$II series (unitary, orthogonal, and symplectic,
respectively) \cite{Dyson}. Three more classes ($A$III, $BD$I, and
$C$II) appear in systems with massless Dirac fermions, as a
consequence of the chiral symmetry \cite{chiral}. Finally,
four more classes ($C$, $D$, $C$I, and $D$III) were
shown to describe mesoscopic superconducting systems, depending
on the presence of the spin-rotation and time-reversal symmetries
\cite{Zirnbauer}. In the present paper we demonstrate 
that the remaining eleventh class $B$ appears in $p$-wave
superconductors under topologically-nontrivial (vortex-type)
boundary conditions responsible for the zero-energy level.

Let us start with briefly describing the properties of
the ensemble of orthogonal matrices in odd dimensions (class $B$).
The Lie algebra $so(2N+1)$ consists of real antisymmetric
$(2N+1)\times (2N+1)$ matrices. Its dimension is 
${\rm dim} [so(2N+1)] = N(2N+1)$ and its rank ${\rm rk}[so(2N+1)]=N$.
If a matrix $A$ belongs to $so(2N+1)$, the Hermitian matrix
$H=iA$ has one zero eigenvalue, and the remaining eigenvalues form
pairs $(\omega_i, -\omega_i)$, $i=1,\dots,N$. 
At low energies we may, without loss of generality, assume the
Gaussian probability distribution for the Hamiltonian
$dP(H)=\exp(-{\rm Tr}H^\dagger H/2v^2) \prod dH_{ij}$, 
where $v$ is a large cut-off energy (in our problem, $v$ is of the
order of the superconducting gap). Then
the joint probability distribution for the (positive) 
eigenvalues $\omega_i$ is of
the conventional form \cite{Zirnbauer}:
\begin{equation}
dP\{\omega_i\}=|J\{\omega_i\}| 
\prod_{i=1}^N e^{-\omega_i^2/v^2} d\omega_i,
\end{equation}
where $J\{\omega_i\}$ is the Jacobian of the diagonalization
of the Hamiltonian,
\begin{equation}
|J\{\omega_i\}| = \prod_{i<j} |\omega_i^2-\omega_j^2|^\beta
\prod_{i=1}^N |\omega_i|^\alpha. 
\label{probability-1}
\end{equation}
At energies much less than the cut-off energy $v$, the 
correlations of the quasiparticle levels $\omega_i$ are
determined solely by the Jacobian $J\{\omega_i\}$
[the expression (\ref{probability-1}) follows from
the explicit form of the roots $\mbox{\boldmath$\xi$}_{(k)}$ of the Lie
algebra $so(2N+1)$ and from
$|J\{\omega_i\}|=\prod_k | \sum_i \xi_{(k)}^i \omega_i |$;
the values of $\alpha$ and $\beta$ may also be
easily found from dimension counting].
For $so(2N+1)$ the parameters of the level
statistics are $\beta=2$, $\alpha=2$. 
The values of $\alpha$ and $\beta$ for the universality class $B$ 
coincide with those for class $C$, and only the 
presence or absence of the
zero-energy level distinguishes the level distributions in the
two classes.

Next, we shall prove that a single-quantum vortex in a two-dimensional
$p$-wave superconductor obeys the statistics of the $so(2N+1)$ ensemble,
provided the symmetry of the Hamiltonian preserves the zero-energy
level. Consider the Bogoliubov-de-Gennes Hamiltonian
\begin{equation}
H=\sum_\alpha \Psi^\dagger_\alpha \left[{({\bf p} -e {\bf A})^2 \over 2m}
+V({\bf r})-\varepsilon_F\right] \Psi_\alpha
+ \Psi_\uparrow^\dagger \left( \Delta_x * {p_x\over k_F} +
\Delta_y * {p_y\over k_F} \right)
\Psi_\downarrow^\dagger + {\rm h.c.},
\label{Hamiltonian}
\end{equation}
where $\Psi_\alpha$ are the electron operators ($\alpha$ is the spin
index), $V({\bf r})$ is the external potential of impurities, 
${\bf A}({\bf r})$ is
the electromagnetic vector potential, $\Delta_x({\bf r})$ and 
$\Delta_y({\bf r})$ are
the coordinate-dependent components of the superconducting gap.
[In the bulk, the preferred superconducting order is one of the two
chiral components $\eta_\pm=\Delta_x\pm\Delta_y$, but in
inhomogeneuos systems, such as a vortex core, an admixture
of the opposite component is self-consistently generated
\cite{p-}. We account for this effect by allowing the
two independent order parameters $\Delta_x$ and $\Delta_y$.]
Star ($*$) denotes the symmetrized ordering of the gradients $p_\mu$ and the
order parameters $\Delta_\mu$ [definition: $A*B\equiv (AB+BA)/2$]. 
At infinity, the order parameters
impose the vortex boundary conditions:
\begin{equation}
\Delta_x(r\to\infty,\phi)=\Delta_0 e^{\pm i\phi}, \qquad
\Delta_y(r\to\infty,\phi)=i\Delta_0 e^{\pm i\phi},
\label{boundary-conditions}
\end{equation}
where $r$ and $\phi$ are polar coordinates.
Plus or minus signs in the exponent correspond to a positive
or a negative single-quantum vortex.
For an axially-symmetric vortex with the chirality of the
order parameter non-self-consistently fixed ($\Delta_y \equiv i \Delta_x$), 
without the vector-potential ${\bf A}({\bf r})$ and without disorder 
$V({\bf r})$,
the low-lying eigenstates of the Hamiltonian (\ref{Hamiltonian})
has been found by Kopnin and Salomaa \cite{Kopnin}. 
The spectrum is 
\begin{equation}
E_n=n \omega_0, \quad (p-{\rm wave}), \qquad n=0,\pm 1, \pm 2, \dots
\label{p-spectrum}
\end{equation}
with $\omega_0 \sim \Delta^2/\varepsilon_F$.
This result should be compared with the spectrum of the
vortex core in a $s$-wave superconductor \cite{deGennes}: 
\begin{equation}
E_n=\left(n+{1\over 2}\right) \omega_0, \quad (s-{\rm wave}),
\qquad n=0,\pm 1, \pm 2, \dots.
\label{s-spectrum}
\end{equation}

The common feature of the spectra in the $s$-wave and $p$-wave cases
is the symmetry about zero energy. If we interpret holes in the
negative-energy levels as excitations with positive energies
(and with the opposite spin), then this symmetry implies that
the excitations are doubly degenerate in spin: to each spin-up
excitation there corresponds a spin-down excitation at the same energy.
For a $s$-wave vortex, this degeneracy is due to the full spin-rotation
$SU(2)$ symmetry. The $p$-wave Hamiltonian (\ref{Hamiltonian})
has a reduced spin symmetry. Namely, it has the symmetry group
$O(2)$ generated by rotations about the $z$-axis 
($\Psi_\uparrow \mapsto e^{i\alpha}\Psi_\uparrow$, 
$\Psi_\downarrow \mapsto e^{-i\alpha}\Psi_\downarrow$) and by the
spin flip $\Psi_\uparrow \mapsto \Psi_\downarrow$, $\Psi_\downarrow
\mapsto \Psi_\uparrow$. This non-abelian group causes the
two-fold degeneracy of all levels (except for the zero-energy
level(s) where the symmetry $O(2)$ may mix the creation
and annihilation operators for the same state). This symmetry
is crucial for our discussion. 
Note that we have not included in the Hamiltonian neither the spin-orbit
term $({\bf U}_{SO} \cdot [\mbox{\boldmath$\sigma$} \times {\bf p}])$,
nor the Zeeman splitting ${\bf H}({\bf r}) \cdot \mbox{\boldmath$\sigma$}$. 
Either of these terms would break the spin
symmetry $O(2)$, which would eventually result in a different universality
class of the disordered system (type $D$ with non-degenerate levels),
if these terms are sufficiently strong.

The difference between the $s$- and $p$-wave vortices is the zero-energy 
level in the $p$-wave case. It has been shown by Volovik that this
level has a topological nature \cite{Volovik}. Indeed, suppose we
gradually increase disorder in the Hamiltonian (\ref{Hamiltonian}).
The levels shift and mix, but the degeneracy of the levels remains
the same as long as the symmetry $O(2)$ is preserved. The total
number of levels remains {\it odd}, and therefore the zero-energy
level {\it cannot shift} if the final Hamiltonian is a continuous
deformation of the original one (without disorder), i.e. if the
topological class of the boundary conditions (\ref{boundary-conditions})
remains the same.

Now we proceed along the usual lines of the random-matrix approach.
Let us take the most random distribution of Hamiltonians within
a given symmetry class. The only symmetry of the
Hamiltonian (\ref{Hamiltonian}) is the spin symmetry $O(2)$. 
The time-reversal symmetry is already broken by the vortex
and by the pairing, and therefore neither
the vector potential ${\bf A}(x)$ nor local deformations of $\Delta_\mu$
can reduce the symmetry of the Hamiltonian. When projected onto
spin-up excitations $\gamma^\dagger_\uparrow = 
\int[u({\bf r})\Psi^\dagger_\uparrow({\bf r}) 
+ v({\bf r})\Psi_\downarrow({\bf r})] d^2 {\bf r}$,
the Hamiltonian for the two-component vector $(u,v)$ takes the form:
\begin{equation}
H=\pmatrix{
\left[ {(-i\nabla - e {\bf A})^2 \over 2m} + V({\bf r}) 
-\varepsilon_F \right] &
\left[{\Delta_x\over k_F}*(-i\nabla_x)+{\Delta_y\over k_F}*(-i\nabla_y)\right]
\cr
\left[{\Delta^*_x\over k_F}*(-i\nabla_x)+
{\Delta^*_y\over k_F}*(-i\nabla_y)\right] &
- \left[ {(-i\nabla + e {\bf A})^2 \over 2m} + V({\bf r}) 
-\varepsilon_F \right] }.
\end{equation}
In an arbitrary orthonormal basis of electronic states, this Hamiltonian
may be written as a matrix
\begin{equation}
H=\pmatrix{
h & \Delta \cr
\Delta^\dagger & -h^*}.
\label{Ham-short}
\end{equation}
From the hermiticity of the Hamiltonian, it follows that $h^\dagger=h$.
From the explicit form of the $p$-wave pairing, $\Delta = -\Delta^T$
(it is here that the $p$-wave structure of the pairing is important;
for $s$-wave pairing we would have $\Delta = \Delta^T$ instead).
These are the only restrictions on the Hamiltonian (\ref{Ham-short}).
If we define 
\begin{equation}
U_0={1\over\sqrt2}\pmatrix{1&1\cr i&-i},
\end{equation}
the restrictions on the Hamiltonian (\ref{Ham-short})
are equivalent to the condition that the rotated matrix
$i U_0 H U_0^{-1}$ is real antisymmetric, i.e. it belongs to the
Lie algebra $so(M)$, where $M$ is the dimension of the 
Hilbert space (the same rotation of the Hamiltonian was used
in Ref.~\cite{Zirnbauer} to identify the $D$ universality class).

The last step in our argument is to note that, under the vortex boundary
conditions, the dimension of the Hamiltonian (\ref{Ham-short})
is odd, not even (this may be difficult to visualize from the
particle-hole representation (\ref{Ham-short}), but easier
from the rotated Hamiltonian $U_0 H U_0^{-1}$). Thus for a single-quantum
vortex, we identify the space of the Hamiltonians with $so(2N+1)$.

A simple consequence of this result is the level distribution
(\ref{probability-1}) with $\alpha=\beta=2$. Besides the zero-energy
level, this distribution is identical to that of class $C$ realized
in $s$-wave vortices \cite{C-ensemble,C-ensemble-2}. This allows us to use the
trick of mapping onto free fermions to compute any correlation
function of the density of states (DOS) \cite{Zirnbauer}. In particular,
the average DOS is
\begin{equation}
\langle \rho(\omega)\rangle ={1\over\omega_0} - {\sin(2\pi\omega / \omega_0) 
\over 2 \pi\omega}
+\delta(\omega),
\end{equation}
where $\omega_0$ is the average inter-level distance. 

One more lesson from our analysis is the difference in the universality
classes of $p$-wave mesoscopic systems from their $s$-wave analogues.
In particular, a $p$-wave system {\it without} a topological
zero-energy level would belong to the universality class $D$
($\beta=2$, $\alpha=0$), in contrast to the class $C$ for its
$s$-wave counterpart. A detectable physical consequence of such
a difference is an increase of the average DOS
near the Fermi energy (in class $D$) as opposed to a suppression
of DOS near Fermi level in class $C$ [in class $B$, such a
suppression is compensated by a $\delta$-function at zero].

In the present work we do not discuss the microscopic derivation
of the level mixing. We assume ``strong level mixing''
which allows us to use the random-matrix approach. On the
other hand, the symmetry of the $p$-wave pairing is known to 
provide certain ``spectrum rigidity'' suppressing shift
and mixing of the low-lying levels by impurities \cite{Volovik}.
Thus to drive the system into the regime of ``strong level
mixing'' may require a stronger disorder than in a $s$-wave vortex.
A microsopic picture of the crossover to the
disordered regime in a $s$-wave case was developed in
\cite{C-ensemble,C-ensemble-2,Larkin}, and its extension to the $p$-wave
case will be a subject of future studies. 
Furthermore, a disorder is known to suppress the $p$-wave
superconductivity, and the possibility to reach the required
level mixing before destroying superconductivity is not obvious.
A coexistence of sufficiently strong disorder and $p$-wave superconductivity
may possibly be achieved in alternative setups such as
disordered normal-superconducting ($p$-wave) hybrid devices.
Besides strong level mixing, the only requirements on the system
to exhibit type $B$ level statistics are the $p$-wave symmetry 
of the pairing and the topological zero-energy level. 

One more approximation made in our model is
neglecting the Zeeman splitting. 
This is a good approximation for strong type II
superconductors (with $\kappa=\lambda / \xi \gg 1$). In
Sr$_2$RuO$_4$ the experiments indicate $\kappa\sim 2.6$ \cite{Ru-exp},
which implies that the Zeeman level shift is of the order
of the inter-level spacing $\omega_0$. However, in a clean vortex,
this shift is approximately constant for the low-lying levels,
due to the large coherence length ($k_F\xi\gg 1$) and to a smooth
magnetic-field profile ${\bf H}({\bf r})$. It is likely that
this property will hold even in the limit of strong disorder, 
then the overall level distribution will be simply shifted 
by a constant energy.

An important observation related to the Zeeman splitting of the
energy levels is the {\it fractional spin} $1/4$ of the vortex
in the ground state. Indeed, in the absence of the Zeeman field,
the multi-particle ground state of the vortex is doubly degenerate,
with the $z$-component of the spin in the two degenerate
states differing by $1/2$. The Zeeman field splits the ground state
as if the vortex were a particle with the $z$-component of the spin
equal {\it half} the spin of the electron. Thus we conclude that
the vortex has the $z$-component of the spin $S_z=\pm 1/4$.
It would be interesting to understand possible physcial implications of
this effect.

Finally, for the convenince of the reader, we have gathered
the information about all eleven symmetry classes in Table 1.
This table is compiled from Refs.\cite{Helgason,Dyson,chiral,Zirnbauer}
and contains the dimensions and ranks of the symmetric
spaces as well as the parameters $\alpha$ and $\beta$ of
the joint probability distributions of the energy levels
(they may be computed solely from the ranks and dimensions
by a simple power counting). The present work
provides the example of the physical system belonging to the
class $B$ (the last line of the table). Besides, one more symmetry
subclass has not been studied so far in the context of
mesoscopics: the odd-$N$ subclass of $D$III. In the work
of Altland and Zirnbauer \cite{Zirnbauer}, the even-$N$
subclass of $D$III is represented by a $s$-wave mesoscopic
system with time-reversal symmetry and with broken spin symmetry.
The novel odd-$N$ subclass of $D$III may occur as a topological
modification of that construction or in a $p$-wave superconducting
system {\it without} time-reversal-symmetry breaking, provided
a zero mode is required by topology. Finding a physically
plausible mesoscopic realization of the odd-$N$ $D$III subclass 
is an interesting problem in the framework of the symmetry classification of
mesoscopic superconducting systems.

\bigskip

\newcommand{\rb}[1]{\raisebox{2ex}[-2ex]{#1}}
{
\renewcommand{\arraystretch}{1.5}
\begin{tabular}{|c|c|c|c|c|c|c|}
\hline
\renewcommand{\arraystretch}{1}
\begin{tabular}{c}
  Cartan \\ class \\
\end{tabular} & 
\renewcommand{\arraystretch}{1}
\begin{tabular}{c}
  Symmetric \\ space \\
\end{tabular} &
Dimension &
Rank &
\hspace*{5pt}$\beta$\hspace*{5pt}&
$\alpha$ & Remarks \\ [10pt]
\hline
$A$ [GUE] & $SU(N)$ & $N^2-1$ & $N-1$ & 2 & $-$ & \\
\cline{1-6}
$A$I [GOE] & $SU(N)/SO(N)$ &  $(N-1)(N+2)/2$ & $N-1$ & 1 & $-$ & 
Wigner-Dyson \cite{Dyson} \\
\cline{1-6}
$A$II [GSE] & $SU(2N)/Sp(N)$ & $(N-1)(2N+1)$ & $N-1$ & 4 & $-$ & \\
\hline
$A$III [chGUE] & $SU(p+q)/S(U(p)\times U(q))$ & 
$2pq$ & $p$ & 2 & $1+2(q-p)$ & 
Chiral ensembles \cite{chiral}, $p \le q$ \\
\cline{1-6}
$BD$I [chGOE] & $SO(p+q)/SO(p)\times SO(q)$ & 
$pq$ & $p$ & 1 & $q-p$ & $(q-p)$ = number of \\
\cline{1-6}
$C$II [chGSE] & $Sp(p+q)/Sp(p)\times Sp(q)$ & 
$4pq$ & $p$ & 4 & $3+4(q-p)$ & zero modes\\
\hline
$C$ & $Sp(N)$ & $N(2N+1)$ & $N$ & 2 & 2 & \\
\cline{1-6}
$D$ & $SO(2N)$ & $N(2N-1)$ & $N$ & 2 & 0 & 
Altland-Zirnbauer \cite{Zirnbauer}\\
\cline{1-6}
$C$I & $Sp(N)/U(N)$ & $N(N+1)$ & $N$ & 1 & 1 & \\
\cline{1-6}
&&&&& 1 (even $N$) & \\
\cline{6-7}
\rb{$D$III} & \rb{$SO(2N)/U(N)$} & \rb{$N(N-1)$} & 
\rb{$\left[{N\over 2}\right]$} & \rb{4} & 
5 (odd $N$) & 
\renewcommand{\arraystretch}{1}
\begin{tabular}{c}
  one zero mode\\
{\it (example unkonwn)} \\
\end{tabular}
\\
\hline
$B$ & $SO(2N+1)$ & $N(2N+1)$ & $N$ & 2 & 2 & 
\renewcommand{\arraystretch}{1}
\begin{tabular}{c}
  one zero mode\\
{\it (present work}\\
{\it  and} Ref.\cite{BD}{\it )} \\
\end{tabular}
\\
\hline
\end{tabular}
}

\smallskip

TABLE 1. Symmetric spaces and universality classes of random-matrix
ensembles.

\bigskip

{\bf P.S.} At the final stage of the preparation of this manuscript,
the author has learned about the recent work of Bocquet, Serban,
and Zirnbauer \cite{BD}, where the type-$B$ level statistics
in $p$-wave vortices has been pointed out.

\bigskip

The author thanks M.~V.~Feigel'man for suggesting this problem, 
for fruitful discussions and for helpful comments on the manuscript. 
Useful discussions with  G.~Blatter, V.~Geshkenbein, D.~Gorokhov, 
R.~Heeb, and M.~Zhitomirsky are greatfully acknowledged.
The author thanks Swiss National Foundation for financial support.

\end{document}